\documentclass{aastex631}
\usepackage{hyperref} 
\usepackage{natbib}
\usepackage{textcomp}
\usepackage{gensymb}
\usepackage{xcolor}
\usepackage{amsmath}
\usepackage{comment}
\usepackage{arydshln}
\usepackage{graphics}
\usepackage{booktabs}

\begin{document}
\title{The thickness of galaxy disks from z=5 to 0 probed by JWST}
\author[0000-0001-5258-1466]{Jianhui Lian}
\affiliation{South-Western Institute for Astronomy Research, Yunnan University, Kunming, Yunnan 650091, People’s Republic of China}
\author{Li Luo}
\affiliation{School of Physics and Astronomy,China West Normal University, Nanchong, Sichuan 637009, People’s Republic of China}


\begin{abstract}
Although thick disk is a structure prevalent in local disk galaxies and also present in our home Galaxy, its formation and evolution is still unclear. Whether the thick disk is born thick and/or gradually heated to be thick after formation is under debate. To disentangle these two scenarios, one effective approach is to inspect the thickness of young disk galaxies in the high redshift Universe. 
In this work we study the vertical structure of 191 edge-on galaxies spanning redshift from 0.2 to 5 using JWST NIRCAM imaging observations. 
For each galaxy, we retrieve the vertical surface brightness profile at 1~${R_e}$ and fit a sech$^2$ function {that has been convolved} with the line spread function. The obtained scale height of galaxies at $z>1.5$ show no clear dependence on redshift, with a median value in remarkable agreement with that of the Milky Way's thick disk. This suggests that local thick disks are already thick when they were formed in the early times and secular heating is unlikely the main driver of thick disk formation. For galaxies at $z<1.5$, however, the disk scale height decreases systematically towards lower redshift, with low-redshift galaxies having comparable scale height with that of the Milky Way's thin disk. This cosmic evolution of disk thickness favors an upside-down formation scenario of galaxy disks.   
\end{abstract}

\section{Introduction}
The thick disk was first discovered in external galaxies \citet{burstein1979,tsikoudi1979}, and then also identified in the Milky Way by counting stars in the vertical direction at the solar radius \citep{gilmore1983,juric2008}. Such geometric thick component was later on found to be prevalent in local disk galaxies \citep{yoachim2006,comeron2012,comeron2018}, albeit suspicion that it may be artificial feature caused by the scatter light \citep{dejong2008,sandin2014}. 
With a careful treatment of the extended wings of the point spread function (PSF), \citet{comeron2018} confirmed the prevalence of thick disk in local edge-on galaxies. 

While the thick disk was first discovered in external galaxies, it is more extensively studied in the Milky Way. Early studies based on solar neighborhood observations have shown that the geometric thick disk of the Milky Way hosts stellar populations that are generally old, chemically metal-poor and $\alpha$-enhanced, and kinematically hot \citep[e.g.,][]{fuhrmann1998,bensby2005,reddy2006,lee2011,adibekyan2012,haywood2013,bensby2014}. More recent massive stellar spectroscopic surveys, however, reveal more complexity of the geometric thick disk which may be composed of different stellar populations at different radii \citep{martig2016,nidever2014,hayden2015,lian2020b,queiroz2020,lian2022}. In particular, the thick disk at larger radii tend to contain stars that are relative younger and less $\alpha$-enhanced. For the thick disks in local galaxies, their stellar population properties have not been extensively studied yet. Based on integral field unit observations of {a few nearby edge-on} galaxies, the local extra-galactic thick disks are likely similar to that of the inner Milky Way {in both stellar population and kinematic properties} \citep{comeron2016,comeron2019,pinna2019}.


One popular explanation of the thick disk formation assumes that the thick disk stars observed today were initially formed within a thin disk and then vertically heated to form a thick disk via interactions with disk symmetries (e.g., spiral arms and Giant Molecular Clouds, \citealt{villumsen1985}) and/or galaxy companions \citep{quinn1993,wyse2006,villalobos2008}. In this scenario, stars born earlier have more chance to be heated and therefore are located in a thicker component, predicting a positive correlation between the age and the vertical extension of stellar populations.  
On the contrary, another scenario that was proposed more recently assumes the local thick disks were initially formed to be thick \citep{brook2004,bournaud2009}. This scenario is supported by the fact that galaxies in the early Universe, when the local thick disk stars are expected to form, are much more turbulent that the local galaxies given their large velocity dispersion \citep[e.g.,][]{weiner2006,shapiro2008,simons2016,ubler2019} and disturbed and clumpy morphology \citep[e.g.,][]{elmegreen2005,elmegreen2009}. 

In galaxy simulations, a mixture of results has yet been reported. In some simulations, the stars in a thick disk today are initially born in a thin component which is then gradually thickened by vertical heating after birth \citep{meng2021,yi2023}. In other simulations \citep{clarke2019,bird2021}, however, the stars already form a thick disk in the early times, but are also then vertically heated afterwards to form a thicker disk. Whether the local thick disks are born already thick or heated to be thick after formation is still under debate.  

While both scenarios can broadly explain the old nature of the thick disks in the local galaxies and the Milky Way, the composite stellar populations of the Milky Way's thick disk revealed by recent massive stellar spectroscopic surveys impose challenges to the heating scenario. In particular, \citet{lian2022} reported the thickness of mono-abundance populations of the Milky Way and found signature of two disconnected phases of thick disk formation separated by a few Gyrs. These results suggest a non-monotonic thickening history of the Milky Way and thus the vertical heating is not likely the major mechanism for the Milky Way's thick disk formation.   

To understand the thick disk formation of galaxies in general, studying the disk thickness of high-redshift edge-on galaxies provides an alternative and intriguing perspective. A few early attempts have been made in this direction using HST observations \citep{elmegreen2006,elmegreen2017,hamilton2023}. However, even with HST spatial resolution, high-redshift edge-on galaxies are marginally resolved. 

In this work, we study the thickness of edge-on galaxies beyong the local Universe using the latest JWST observations, which have higher spatial resolution and are deeper than the HST observations that allow us to study the vertical structure of more distant galaxies. Throughout this paper, we assume the cosmological parameters ($\Omega_{\rm M},\ \Omega_{\Lambda}, h_0$) = (0.27, 0.73, 0.71) and adopt AB magnitude system.  

\section{Data}
\subsection{Sample selection}
In this work, we select edge-on galaxies using public JWST/NIRCAM imaging {data from} Cosmic Evolution Early Release Science (CEERS) survey. CEERS survey (PI: S. Finkelstein) is designed to cover 100 arcmin$^2$ within Extended Groth Strip (EGS) field with JWST infrared imaging and spectroscopy using NIRCam, MIRI, and NIRSpec \citep{finkelstein2023}. We use the JWST imaging data that are reduced with grizli \footnote{10.5281/zenodo.1146904} and msaexp \citep{valentino2023} and released in the DAWN JWST Archive (DJA), which is a repository of public JWST galaxy data. 

We focus on the JWST observations at one single band, F115W, which among all JWST filters has {currently} the longest exposure time in the CEERS survey and is a filter at short wavelength with relatively sharp PSF. In the released F115W images in the CEERS survey, a total number of 98149 sources are detected. For each source, we retrieve a 50$\times$50 pixel ($\sim2^{\prime\prime}\times2^{\prime\prime}$) image cut and obtain basic shape and photometry parameters (e.g., minor-to-major axis ratio, position angle) using {\sc{Sextractor}}. We then measure the effective radii of these galaxies by extracting light profile using a series of ellipses with the same shape parameters derived above. Since the JWST CEERS field overlaps with the HST EGS field, we adopt the physical properties of these galaxies, including photometric/spectroscopic redshifts, stellar mass, and star formation rate (SFR) from public EGS catalogs (i.e., mass and redshift from \citet{stefanon2017} and SFR from \citet{barro2019}).

We then apply the following criteria to select our sample of edge-on galaxies: 
\begin{itemize}
\item edge-on with axes ratio $<0.4$;
\item relatively massive with stellar mass ${\rm log(M_{*}/M_{\odot})>8.5}$;
\item star forming with specific star formation rate ${\rm log(sSFR/yr^{-1})>-11}$;
\item redshift between 0 and 5. 
\end{itemize}
A number of 352 galaxies satisfy these criteria. By visual inspection, we further remove 161 galaxies which are either wings of bright stars, part of nearby spiral galaxies, or close to companion galaxies or bright stars. Finally, the remaining 191 galaxies consist of our edge-on galaxy sample. For three of them, spectroscopic redshift is available. 

\begin{figure*}
	\centering
	\includegraphics[width=0.8\textwidth]{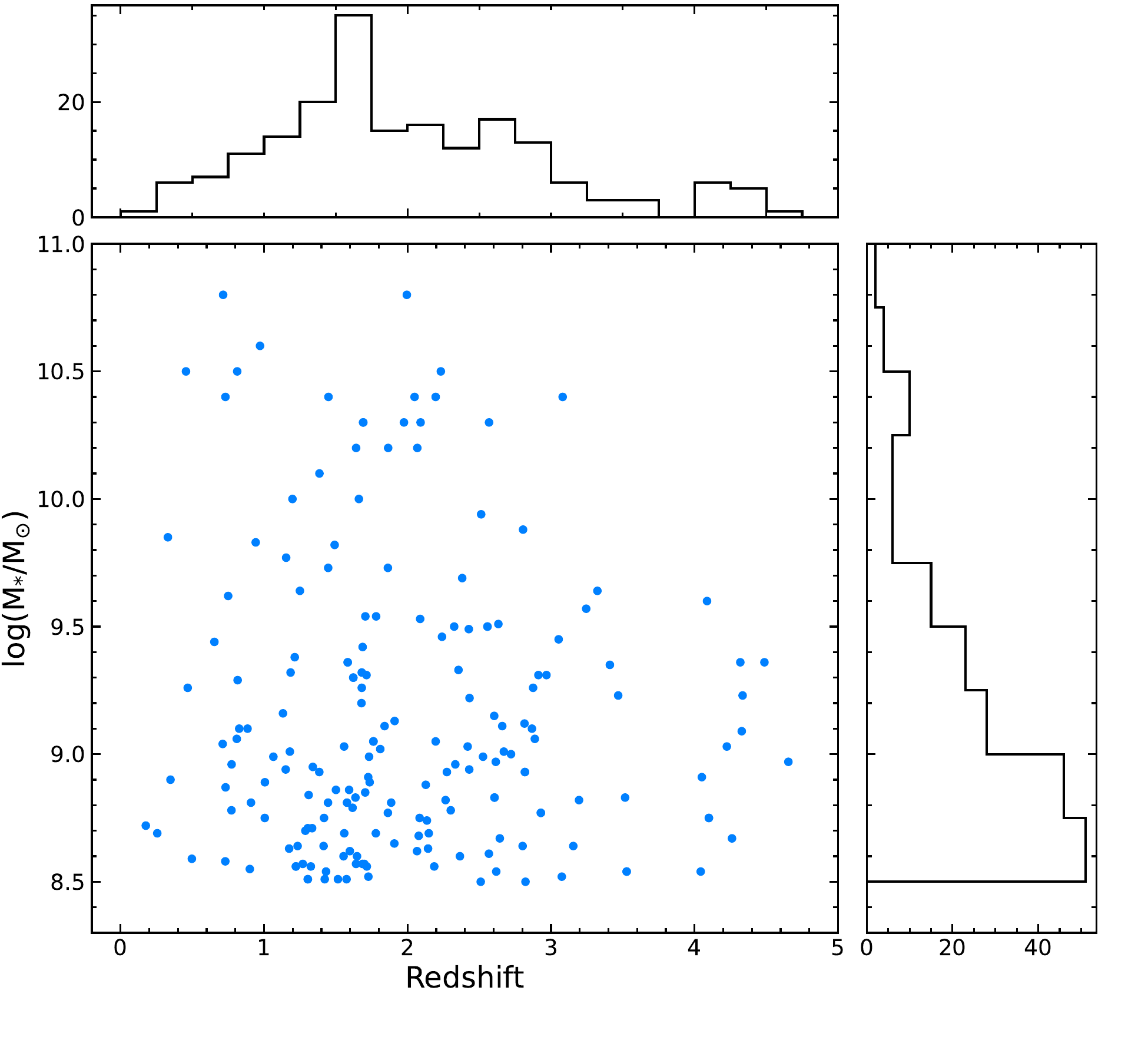}
	\caption{The distribution of stellar mass and redshift of our edge-on galaxy sample. }
	\label{redz-mass}
\end{figure*}

Figure~\ref{redz-mass} shows the distribution of our edge-on galaxy sample in stellar mass and redshift. The majority of the sample have a stellar mass below $10^{9.5}{\rm M_{\odot}}$ and redshift between 1 and 3. 

\subsection{Construction of the line spread function}
\begin{figure*}
	\centering
	\includegraphics[width=0.8\textwidth]{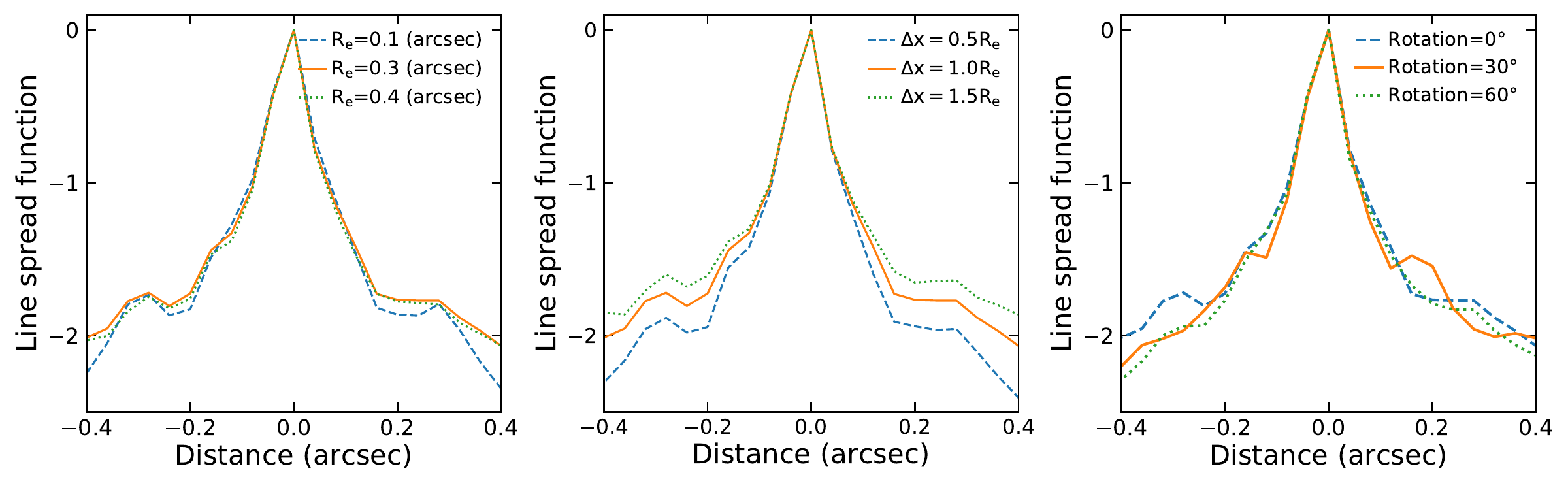}
	\caption{The 1D point spread function (PSF) that is used for modelling the vertical surface brightness profile of edge-on galaxies in our sample. The 1D PSF is constructed by convolving the 2D PSF of JWST F115W images with an infinitesimally thick disk model of a given size. The left panel shows the 1D PSF measured at one effective radii for disk models of different sizes. The right panel illustrates the 1D PSF measured at different position relative to the center for the same disk model with effective radius of 0.2\arcsec. }
	\label{psf}
\end{figure*}

To account for the effect of the instrument broadening on the light distribution in the vertical direction, we derive the line spread function (LSF) following the approach introduced in \citet{elmegreen2017} and \citet{comeron2018}. We first {generate NIRCAM 2D PSF model at F115W band oversampled in a $4\times4$ grid with pixel scale of 0.01$^{\prime\prime}$ using WebbPSF\footnote{https://webbpsf.readthedocs.io/en/stable/index.html} \citep{perrin2014}. We then convolve this PSF model} 
with an infinitely thin edge-on galaxy that has a radially declining profile. {From the mock galaxy image we measure the intensity profile} perpendicular to the mid-plane at {a certain radius} which is assumed to be the line spread function. Figure~\ref{psf} shows the obtained LSFs for input galaxy with various effective radius (${\rm R_e}$,{left} panel), measured at different radial position for an input galaxy with ${\rm R_e=0.2^{\prime\prime}}$ ({middle} panel), {and convolved with the 2D PSF model rotated in different angles (right panel)}. 

It can be seen that the LSFs show no clear dependence on the size of the input galaxy and weak dependence on the radial position where the LSF is measured {and the orientation of the PSF}. In this work we extract the vertical brightness profile of observed galaxies at 1~${\rm R_e}$ to measure the scale height. Therefore we adopt the LSF measured at 1~${\rm R_e}$ with input galaxy ${\rm R_e=0.2^{\prime\prime}}$ {and non-rotated PSF}. {We have tested that the scale height measurements with different PSF orientations vary less than 10\%.} 
The FWHM of the adopted LSF is 0.039\arcsec, a factor of {six} smaller than that of the HST images at similar wavelength range in the near-infrared \citep{hamilton2023}. This demonstrates the advanced ability of JWST to spatially resolve distant galaxies. 

\section{Results and Discussion}
\subsection{Vertical surface brightness profile (VSBP)}
\begin{figure*}
	\centering
	\includegraphics[width=\textwidth]{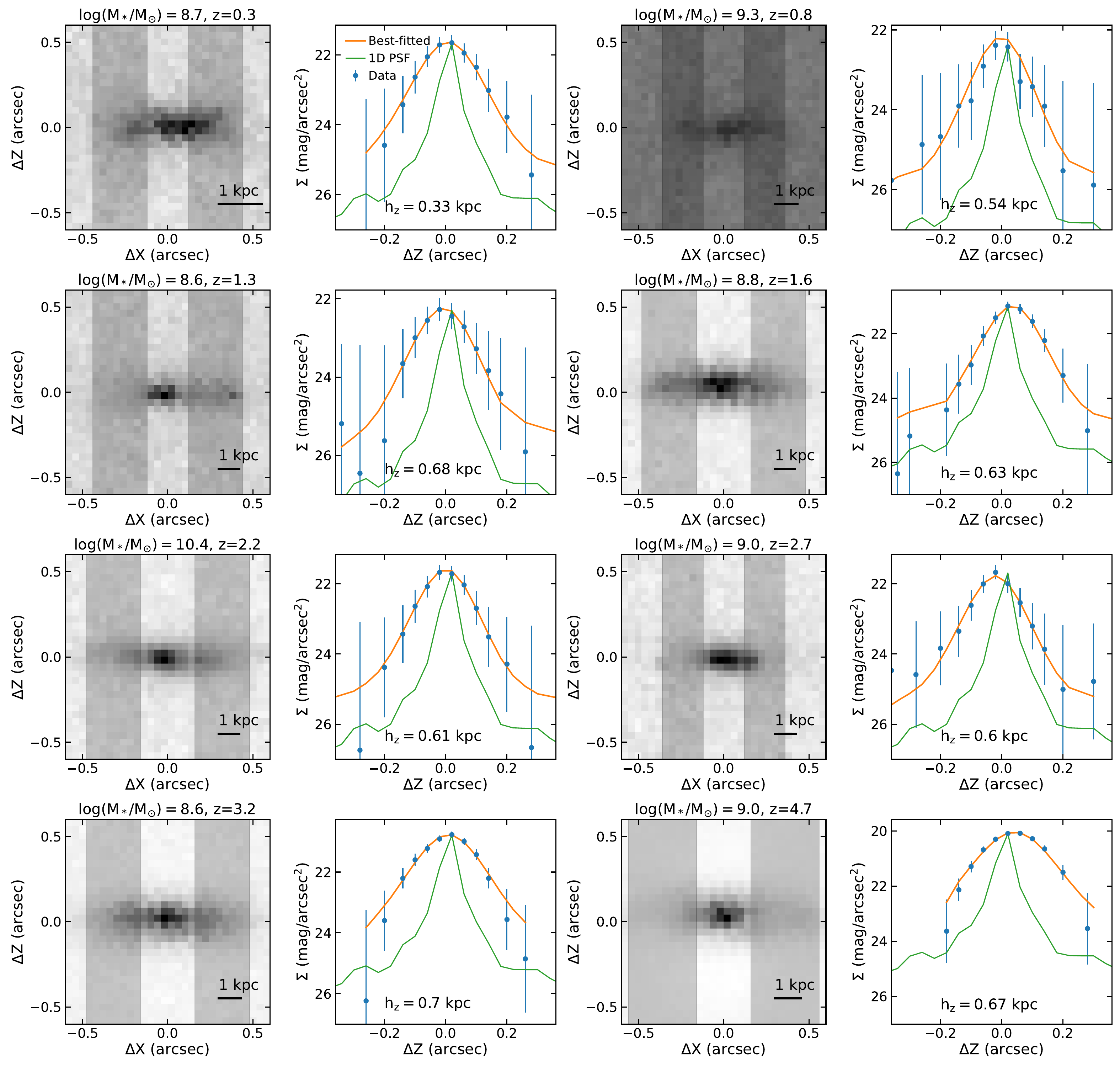}
	\caption{JWST/F115W images and vertical surface brightness profiles of eight example galaxies from low redshift at top left to high redshift at bottom right. In the image, the two gray shaded regions cover the pixels within 0.5$--$1.5~${\rm R_e}$. The length of 1~kpc at each corresponding redshift is illustrated at the bottom right corner of the image. In the VSBP panel, the retrieved VSBP and best-fitted sech$^2$ model are shown as blue points and orange curve. The obtained sech$^2$ scale length is presented at the bottom. The line spread function obtained in \textsection2.2 is also included as reference. }
	\label{example}
\end{figure*}

To measure the disk thickness of our edge-on galaxies, we first retrieve the vertical surface brightness profile (VSBP) at 1~${\rm R_e}$, which is represented by the median profile between 0.5 and 1.5~${\rm R_e}$. The vertical density profiles of galactic disks are well described by sech$^2$ function \citep[e.g.,][]{xiang2018,hamilton2023}. We therefore fit the VSBP of our galaxies with a sech$^2$ function
\begin{equation*}
{\rm \Sigma=\mu_0\times\frac{4}{{(e^{\Delta z/hz}+e^{-\Delta z/hz})}^2}}
\end{equation*}
{that has been convolved} with the LSF obtained in \textsection2.2. Here $\mu_0$ and ${\rm h_z}$ are the central surface brightness and the scale height, respectively. The sech$^2$ function approximates exponential form at large distance, but the scale height in the sech$^2$ function is one-half of the exponential scale height.

Figure~\ref{example} shows the JWST/F115W image and retrieved VSBPs of eight example edge-on galaxies at different redshifts. It can be seen that the observed VSBPs are significantly broader than the LSF, meaning that these galaxies are well resolved by JWST. The best-fitted sech$^2$ function are shown as orange line of which the scale height is shown at the bottom.  

The uncertainties of the scale height measurement is estimated using Mento Carlo simulation. For each galaxy, we { first compute a randomised} VSBP based on the fiducial profile and associated uncertainties {estimated from the weight images that have taken the sky background noise into account. We then redo the fitting for the new VSBP.} After repeating this process 100 times, we take the standard deviation of the 100 measurements as the uncertainty of the scale height measurement. This uncertainty estimate, however, is not available for {65} out of 191 galaxies due to large uncertainties in the VSBP that stop the fitting from converging for the randomized VSBP.  

\subsection{Evolution of disk thickness }
\begin{figure*}
	\centering
	\includegraphics[width=\textwidth]{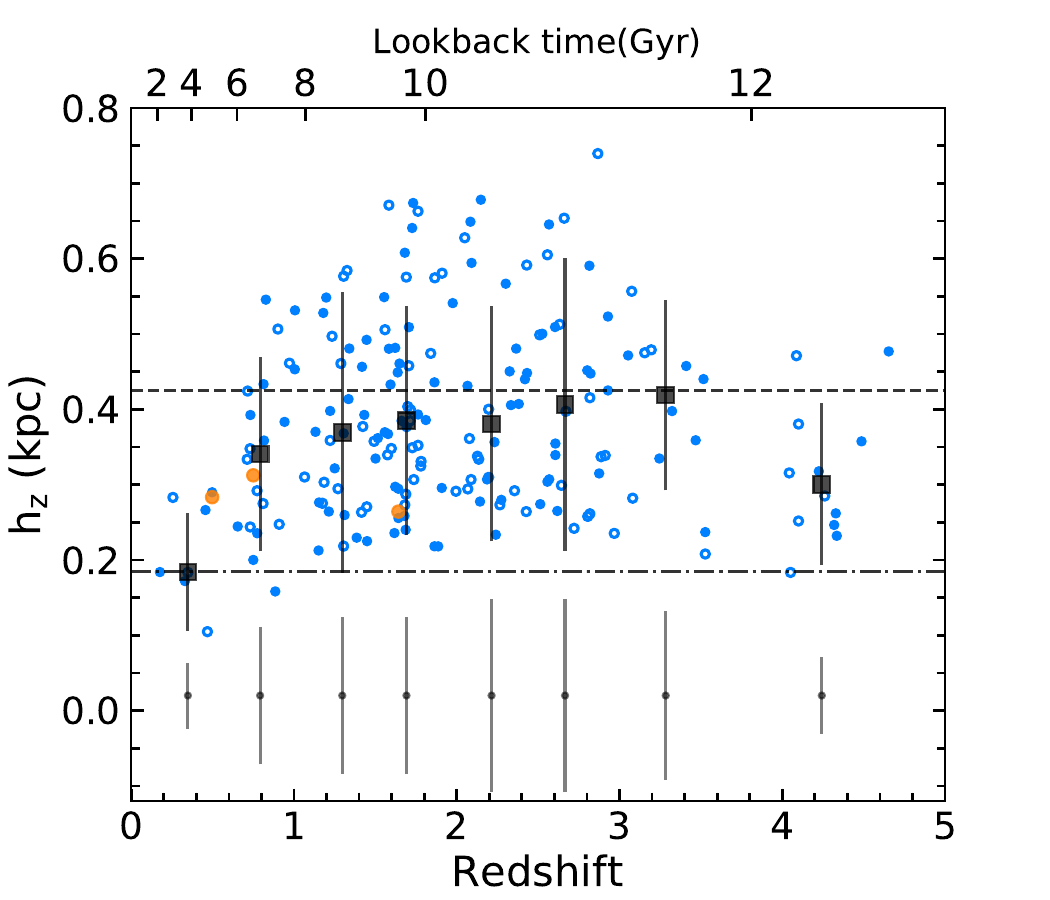}
	\caption{Redshift evolution of galaxy disk thickness. Small blue circles indicate the sech$^2$ scale height measurements of invidividual edge-on galaxies with only photometric redshift estimate and orange circles for those with spectroscopic redshift. Filled circles are the galaxies with valid uncertainty estimate and empty circles without. The level of uncertainties is shown at the bottom in gray. Enlarged black squares and error bars show the median scale length and $1\sigma$ scatter at each redshift bin with bin width of 0.5 at $z<3$ and bin width of 1 at $z>3$. The converted sech$^2$ scale height of the Milky Way's thick and thin disks are denoted by the dashed and dash-dotted lines, respectively.}
	\label{thick-redz}
\end{figure*}

\begin{figure*}
	\centering
	\includegraphics[width=\textwidth]{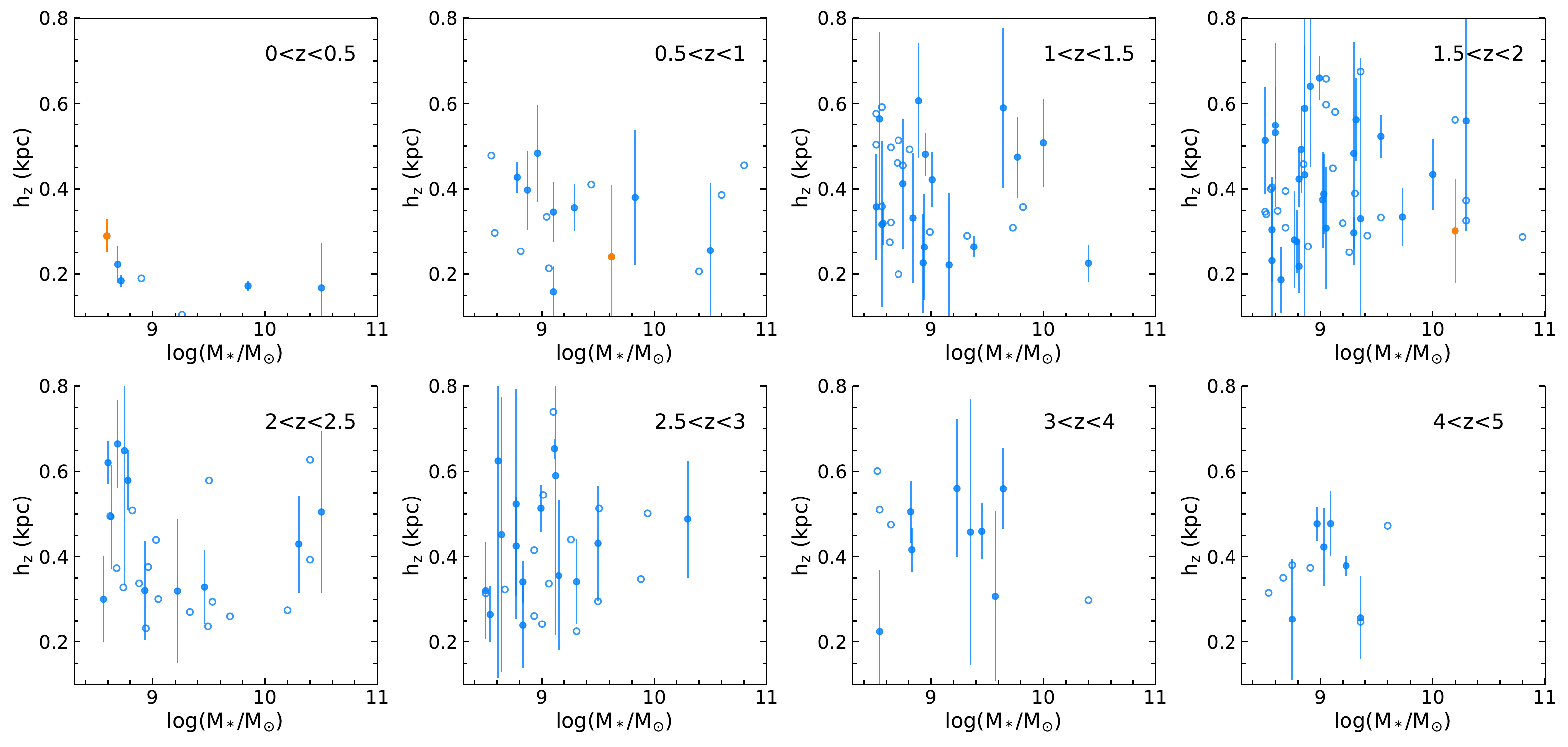}
	\caption{Scale height as a function of stellar mass for galaxies in eight redshift bins. Each panel shows edge-on galaxies in one redshift bin. The symbols are the same as Figure~\ref{thick-redz}.}
	\label{thick-mass}
\end{figure*}

The obtained scale height of our edge-on galaxies as a function of redshift is shown in Figure~\ref{thick-redz}. The enlarged squares indicate the median trend with redshift bin width of 0.5 at $z<3$ and bin width of 1 at $z>3$. As shown in \citet{hamilton2023}, the scale height of galaxies selected with axis ratio $<0.4$ would be on average overestimated by 22\% because of projection effect in case of non-zero inclination to our line-of-sight. This correction factor has been applied to our scale height measurements {as shown in Figure~\ref{thick-redz}}. 
For comparison, we include the scale height measurement of the Milky Way's chemical thick (dashed line) and thin discs (dash-dotted line) from \citet{lian2022}. Since they are derived with an exponential fit, we divide the measurements by a factor of 2 to make them comparable to the sech$^2$ scale height in this work. 

At $z>1.5$, there is a large scatter of scale height among individual edge-on galaxies with no clear trend with redshift. The median sech$^2$ scale height is {0.38$\pm0.13$~kpc}, in remarkable consistence with that of the Milky Way's thick disk (0.43~kpc), suggesting that the Galactic thick disk is already thick in the early times. A similarly large scale height of high-redshift galaxies was also found in previous works using HST observations \citep{elmegreen2006,hamilton2023}. This result suggests that the secular heating is unlikely the main formation mechanism of thick disk. 

Interestingly, at $z<1.5$, the galaxy disk thickness decreases towards lower redshift, reaching a comparable value of the Milky Way's thin disk at low redshift of {$z\sim0.4$}. This trend is significant with Pearson correlation coefficient of {0.36}. However, no such trend was found before in previous relevant studies based on HST observations \citep{elmegreen2006,hamilton2023}. The lack of redshift evolution of disk thickness in these studies is possibly because their distant edge-on galaxies are marginally resolved given the much broader PSF of HST. However, it is worth noting that our edge-on galaxy sample at $z<1$ is still small. We expect this could be greatly improved by {future} wider survey of JWST and forthcoming ground- and space-based wide-field imaging surveys in the near future (e.g., LSST).   


The observed redshift evolution of the galaxy disk thickness since $z\sim1.5$ strongly favors an upside-down formation scenario of galactic disks. In this scenario, the early galaxy disk at high redshift is initially born 
thick because of high turbulence in the disk that may be caused by, for example, frequent galaxy mergers and interactions, {accretion of gas, and star formation feedback. Gas turbulence makes the gas disk thick and therefore the star formation and young stellar disk thick.} 
As galaxies grow and evolve to the low redshift Universe, the gravitational potential well become deeper and merger events become less frequent. As a result, the disk turbulence decreases and star formation occurs in an increasingly thinner disk.   
This scenario has also been proposed to explain the thick disk formation and evolution of the Milky Way that is revealed in great detail with massive spectroscopic stellar surveys \citep{bovy2012b,bovy2016,mackereth2017,yu2021,lian2022}. 

In contrast to upside-down formation scenario, the secular heating scenario suggests the galaxy disk to be born thin and heated to be thick gradually. These are the opposite of the results shown in Fig.~\ref{thick-redz} and therefore the heating scenario is unlikely the main driver behind the thick disk formation. To further investigate the role played by secular heating in thick disk formation and evolution, one need to decompose the thin and thick disk of low-redshift galaxies and connect the thick disk of galaxies at different redshifts. 


To inspect the mass dependence of galaxy disk thickness, we plot the scale height of our edge-on galaxies against their stellar mass in Figure~\ref{thick-mass}. The sample is split into eight redshift bins to minimize the variation of disk thickness due to redshift. For galaxies at {$z>1$}, there is a tentative positive correlation between the disk thickness and stellar mass. {This mass dependence is not clearly seen at low redshift.} Different from our result, \citet{elmegreen2017} found a strong positive correlation between the stellar mass and disk thickness for a sample of 107 edge-on galaxies at $z<1.5$ using HST ACS observations. 

The measurements of disk scale height for all of our galaxies reported here are based on imaging data in a single band of F115W. Therefore for galaxies at different redshifts, the scale height are actually measured in different restframe wavelength range that may trace stellar populations of different ages. \citet{elmegreen2006} found that, for the same galaxy, the sech$^2$ scale height measured based on observations at longer wavelength is systematically higher. To investigate whether and how our results are affected by the choice of the filter for the imaging data, we have measured the scale heights of our galaxies using JWST imaging data in longer wavelength filters, including F150W, F200W, F277W, F356W, and F444W following the same procedure as we adopt in F115W. We find that the obtained individual scale height measurements and redshift trend are remarkably consistent with that obtained in F115W band, especially for galaxies at $z>1$. This is understandable because high-redshift galaxies are generally young and thus the stellar populations traced by observations in different wavelength range are not in large difference. We conclude that our results are not dependent on the choice of the filter used for the scale height measurement.  

{To verify that our measurement of disk thickness are not significantly contaminated by the bulge structure, we also extract the VSBP at larger radii between 0.8-1.8~${R_e}$ and repeat the analysis above. The obtained scale heights are in good agreement with the measurements presented above, suggesting that our results are not significantly affected by the possible existence of bulge structure.} 

\section{Conclusion}
In this work we study the thickness of edge-on disk galaxies beyond the local Universe using high quality imaging observation {of} JWST in the CEERS field. With a much sharper PSF than HST, the vertical structure of {distant} edge-on galaxies are better resolved by JWST. 

Based on a sample of 191 edge-on galaxies with redshift $0.2<z<5$ and stellar mass ${\rm 8.5<log(M_*/M_{\odot})<11}$, we find high-redshift galaxies at $z>1.5$ are generally thicker than the galaxies at lower redshift, with a scale height comparable to that of the Milky Way's thick disk. This suggests that the thick disks in galaxies and the Milky Way are likely born thick instead of gradually thickening after formation. At $z<1.5$, we find an interesting trend of decreasing disk thickness of galaxies at lower redshift with the thickness of low-redshift galaxies comparable to that of the Milky Way's thin disk. This trend serves as strong evidence for the upside-down formation of galaxy disk. 

Besides the redshift, we also investigate the dependence of disk thickness on stellar mass and find {weak} mass dependence except for galaxies at {$z<1$}. 

Our current edge-on galaxy sample {is still relatively small, especially} at $z<1$. We expect forthcoming wide-field imaging surveys in the future will greatly increase the sample of {intermediate- and high-redshift} edge-on galaxies and facilitate the studies of galaxy structure evolution. 

\section*{Acknowledgements}
JL acknowledges support by Yunnan Province Science and Technology Department under Grant No. 202105AE160021 and No. 202005AB160002 {and by Yunnan University grant No. CY22623101}. JL is grateful to Xiaowei Liu, Guanwen Fang, Jie Song, Cheng Cheng, and Xianzhong Zheng for useful discussions.  
{The data were obtained from the Mikulski Archive for Space Telescopes at the Space Telescope Science Institute, which is operated by the Association of Universities for Research in Astronomy, Inc., under NASA contract NAS 5–03127 for JWST. The specific observations analyzed can be accessed via doi:10.17909/g3nt-a370. The high-level imaging data} products were retrieved from the Dawn JWST Archive (DJA). DJA is an initiative of the Cosmic Dawn Center, which is funded by the Danish National Research Foundation under grant No. 140.
\bibliographystyle{aasjournal}
\bibliography{Jianhui}{}

\begin{thebibliography}{}
\expandafter\ifx\csname natexlab\endcsname\relax\def\natexlab#1{#1}\fi
\providecommand{\url}[1]{\href{#1}{#1}}
\providecommand{\dodoi}[1]{doi:~\href{http://doi.org/#1}{\nolinkurl{#1}}}
\providecommand{\doeprint}[1]{\href{http://ascl.net/#1}{\nolinkurl{http://ascl.net/#1}}}
\providecommand{\doarXiv}[1]{\href{https://arxiv.org/abs/#1}{\nolinkurl{https://arxiv.org/abs/#1}}}

\bibitem[{{Adibekyan} {et~al.}(2012){Adibekyan}, {Sousa}, {Santos}, {Delgado
  Mena}, {Gonz{\'a}lez Hern{\'a}ndez}, {Israelian}, {Mayor}, \&
  {Khachatryan}}]{adibekyan2012}
{Adibekyan}, V.~Z., {Sousa}, S.~G., {Santos}, N.~C., {et~al.} 2012, \aap, 545,
  A32, \dodoi{10.1051/0004-6361/201219401}

\bibitem[{{Barro} {et~al.}(2019){Barro}, {P{\'e}rez-Gonz{\'a}lez}, {Cava},
  {Brammer}, {Pandya}, {Eliche Moral}, {Esquej}, {Dom{\'\i}nguez-S{\'a}nchez},
  {Alcalde Pampliega}, {Guo}, {Koekemoer}, {Trump}, {Ashby}, {Cardiel},
  {Castellano}, {Conselice}, {Dickinson}, {Dolch}, {Donley}, {Espino Briones},
  {Faber}, {Fazio}, {Ferguson}, {Finkelstein}, {Fontana}, {Galametz},
  {Gardner}, {Gawiser}, {Giavalisco}, {Grazian}, {Grogin}, {Hathi}, {Hemmati},
  {Hern{\'a}n-Caballero}, {Kocevski}, {Koo}, {Kodra}, {Lee}, {Lin}, {Lucas},
  {Mobasher}, {McGrath}, {Nandra}, {Nayyeri}, {Newman}, {Pforr}, {Peth},
  {Rafelski}, {Rodr{\'\i}guez-Munoz}, {Salvato}, {Stefanon}, {van der Wel},
  {Willner}, {Wiklind}, \& {Wuyts}}]{barro2019}
{Barro}, G., {P{\'e}rez-Gonz{\'a}lez}, P.~G., {Cava}, A., {et~al.} 2019, \apjs,
  243, 22, \dodoi{10.3847/1538-4365/ab23f2}

\bibitem[{{Bensby} {et~al.}(2005){Bensby}, {Feltzing}, {Lundstr{\"o}m}, \&
  {Ilyin}}]{bensby2005}
{Bensby}, T., {Feltzing}, S., {Lundstr{\"o}m}, I., \& {Ilyin}, I. 2005, \aap,
  433, 185, \dodoi{10.1051/0004-6361:20040332}

\bibitem[{{Bensby} {et~al.}(2014){Bensby}, {Feltzing}, \& {Oey}}]{bensby2014}
{Bensby}, T., {Feltzing}, S., \& {Oey}, M.~S. 2014, \aap, 562, A71,
  \dodoi{10.1051/0004-6361/201322631}

\bibitem[{{Bird} {et~al.}(2021){Bird}, {Loebman}, {Weinberg}, {Brooks},
  {Quinn}, \& {Christensen}}]{bird2021}
{Bird}, J.~C., {Loebman}, S.~R., {Weinberg}, D.~H., {et~al.} 2021, \mnras, 503,
  1815, \dodoi{10.1093/mnras/stab289}

\bibitem[{{Bournaud} {et~al.}(2009){Bournaud}, {Elmegreen}, \&
  {Martig}}]{bournaud2009}
{Bournaud}, F., {Elmegreen}, B.~G., \& {Martig}, M. 2009, \apjl, 707, L1,
  \dodoi{10.1088/0004-637X/707/1/L1}

\bibitem[{{Bovy} {et~al.}(2012){Bovy}, {Rix}, {Liu}, {Hogg}, {Beers}, \&
  {Lee}}]{bovy2012b}
{Bovy}, J., {Rix}, H.-W., {Liu}, C., {et~al.} 2012, \apj, 753, 148,
  \dodoi{10.1088/0004-637X/753/2/148}

\bibitem[{{Bovy} {et~al.}(2016){Bovy}, {Rix}, {Schlafly}, {Nidever},
  {Holtzman}, {Shetrone}, \& {Beers}}]{bovy2016}
{Bovy}, J., {Rix}, H.-W., {Schlafly}, E.~F., {et~al.} 2016, \apj, 823, 30,
  \dodoi{10.3847/0004-637X/823/1/30}

\bibitem[{{Brook} {et~al.}(2004){Brook}, {Kawata}, {Gibson}, \&
  {Freeman}}]{brook2004}
{Brook}, C.~B., {Kawata}, D., {Gibson}, B.~K., \& {Freeman}, K.~C. 2004, \apj,
  612, 894, \dodoi{10.1086/422709}

\bibitem[{{Burstein}(1979)}]{burstein1979}
{Burstein}, D. 1979, \apj, 234, 829, \dodoi{10.1086/157563}

\bibitem[{{Clarke} {et~al.}(2019){Clarke}, {Debattista}, {Nidever}, {Loebman},
  {Simons}, {Kassin}, {Du}, {Ness}, {Fisher}, {Quinn}, {Wadsley}, {Freeman}, \&
  {Popescu}}]{clarke2019}
{Clarke}, A.~J., {Debattista}, V.~P., {Nidever}, D.~L., {et~al.} 2019, \mnras,
  484, 3476, \dodoi{10.1093/mnras/stz104}

\bibitem[{{Comer{\'o}n} {et~al.}(2018){Comer{\'o}n}, {Salo}, \&
  {Knapen}}]{comeron2018}
{Comer{\'o}n}, S., {Salo}, H., \& {Knapen}, J.~H. 2018, \aap, 610, A5,
  \dodoi{10.1051/0004-6361/201731415}

\bibitem[{{Comer{\'o}n} {et~al.}(2019){Comer{\'o}n}, {Salo}, {Knapen}, \&
  {Peletier}}]{comeron2019}
{Comer{\'o}n}, S., {Salo}, H., {Knapen}, J.~H., \& {Peletier}, R.~F. 2019,
  \aap, 623, A89, \dodoi{10.1051/0004-6361/201833653}

\bibitem[{{Comer{\'o}n} {et~al.}(2016){Comer{\'o}n}, {Salo}, {Peletier}, \&
  {Mentz}}]{comeron2016}
{Comer{\'o}n}, S., {Salo}, H., {Peletier}, R.~F., \& {Mentz}, J. 2016, \aap,
  593, L6, \dodoi{10.1051/0004-6361/201629292}

\bibitem[{{Comer{\'o}n} {et~al.}(2012){Comer{\'o}n}, {Elmegreen}, {Salo},
  {Laurikainen}, {Athanassoula}, {Bosma}, {Knapen}, {Gadotti}, {Sheth}, {Hinz},
  {Regan}, {Gil de Paz}, {Mu{\~n}oz-Mateos}, {Men{\'e}ndez-Delmestre},
  {Seibert}, {Kim}, {Mizusawa}, {Laine}, {Ho}, \& {Holwerda}}]{comeron2012}
{Comer{\'o}n}, S., {Elmegreen}, B.~G., {Salo}, H., {et~al.} 2012, \apj, 759,
  98, \dodoi{10.1088/0004-637X/759/2/98}

\bibitem[{{de Jong}(2008)}]{dejong2008}
{de Jong}, R.~S. 2008, \mnras, 388, 1521,
  \dodoi{10.1111/j.1365-2966.2008.13505.x}

\bibitem[{{Elmegreen} \& {Elmegreen}(2006)}]{elmegreen2006}
{Elmegreen}, B.~G., \& {Elmegreen}, D.~M. 2006, \apj, 650, 644,
  \dodoi{10.1086/507578}

\bibitem[{{Elmegreen} {et~al.}(2017){Elmegreen}, {Elmegreen}, {Tompkins}, \&
  {Jenks}}]{elmegreen2017}
{Elmegreen}, B.~G., {Elmegreen}, D.~M., {Tompkins}, B., \& {Jenks}, L.~G. 2017,
  \apj, 847, 14, \dodoi{10.3847/1538-4357/aa88d4}

\bibitem[{{Elmegreen} {et~al.}(2009){Elmegreen}, {Elmegreen}, {Marcus},
  {Shahinyan}, {Yau}, \& {Petersen}}]{elmegreen2009}
{Elmegreen}, D.~M., {Elmegreen}, B.~G., {Marcus}, M.~T., {et~al.} 2009, \apj,
  701, 306, \dodoi{10.1088/0004-637X/701/1/306}

\bibitem[{{Elmegreen} {et~al.}(2005){Elmegreen}, {Elmegreen}, {Rubin}, \&
  {Schaffer}}]{elmegreen2005}
{Elmegreen}, D.~M., {Elmegreen}, B.~G., {Rubin}, D.~S., \& {Schaffer}, M.~A.
  2005, \apj, 631, 85, \dodoi{10.1086/432502}

\bibitem[{{Finkelstein} {et~al.}(2023){Finkelstein}, {Bagley}, {Ferguson},
  {Wilkins}, {Kartaltepe}, {Papovich}, {Yung}, {Arrabal Haro}, {Behroozi},
  {Dickinson}, {Kocevski}, {Koekemoer}, {Larson}, {Le Bail}, {Morales},
  {P{\'e}rez-Gonz{\'a}lez}, {Burgarella}, {Dav{\'e}}, {Hirschmann},
  {Somerville}, {Wuyts}, {Bromm}, {Casey}, {Fontana}, {Fujimoto}, {Gardner},
  {Giavalisco}, {Grazian}, {Grogin}, {Hathi}, {Hutchison}, {Jha}, {Jogee},
  {Kewley}, {Kirkpatrick}, {Long}, {Lotz}, {Pentericci}, {Pierel}, {Pirzkal},
  {Ravindranath}, {Ryan}, {Trump}, {Yang}, {Bhatawdekar}, {Bisigello}, {Buat},
  {Calabr{\`o}}, {Castellano}, {Cleri}, {Cooper}, {Croton}, {Daddi}, {Dekel},
  {Elbaz}, {Franco}, {Gawiser}, {Holwerda}, {Huertas-Company}, {Jaskot},
  {Leung}, {Lucas}, {Mobasher}, {Pandya}, {Tacchella}, {Weiner}, \&
  {Zavala}}]{finkelstein2023}
{Finkelstein}, S.~L., {Bagley}, M.~B., {Ferguson}, H.~C., {et~al.} 2023, \apjl,
  946, L13, \dodoi{10.3847/2041-8213/acade4}

\bibitem[{{Fuhrmann}(1998)}]{fuhrmann1998}
{Fuhrmann}, K. 1998, \aap, 338, 161

\bibitem[{{Gilmore} \& {Reid}(1983)}]{gilmore1983}
{Gilmore}, G., \& {Reid}, N. 1983, \mnras, 202, 1025,
  \dodoi{10.1093/mnras/202.4.1025}

\bibitem[{{Hamilton-Campos} {et~al.}(2023){Hamilton-Campos}, {Simons},
  {Peeples}, {Snyder}, \& {Heckman}}]{hamilton2023}
{Hamilton-Campos}, K.~A., {Simons}, R.~C., {Peeples}, M.~S., {Snyder}, G.~F.,
  \& {Heckman}, T.~M. 2023, arXiv e-prints, arXiv:2303.04171,
  \dodoi{10.48550/arXiv.2303.04171}

\bibitem[{{Hayden} {et~al.}(2015){Hayden}, {Bovy}, {Holtzman}, {Nidever},
  {Bird}, {Weinberg}, {Andrews}, {Majewski}, {Allende Prieto}, {Anders},
  {Beers}, {Bizyaev}, {Chiappini}, {Cunha}, {Frinchaboy},
  {Garc{\'\i}a-Her{\'n}and ez}, {Garc{\'\i}a P{\'e}rez}, {Girardi}, {Harding},
  {Hearty}, {Johnson}, {M{\'e}sz{\'a}ros}, {Minchev}, {O'Connell}, {Pan},
  {Robin}, {Schiavon}, {Schneider}, {Schultheis}, {Shetrone}, {Skrutskie},
  {Steinmetz}, {Smith}, {Wilson}, {Zamora}, \& {Zasowski}}]{hayden2015}
{Hayden}, M.~R., {Bovy}, J., {Holtzman}, J.~A., {et~al.} 2015, \apj, 808, 132,
  \dodoi{10.1088/0004-637X/808/2/132}

\bibitem[{{Haywood} {et~al.}(2013){Haywood}, {Di Matteo}, {Lehnert}, {Katz}, \&
  {G{\'o}mez}}]{haywood2013}
{Haywood}, M., {Di Matteo}, P., {Lehnert}, M.~D., {Katz}, D., \& {G{\'o}mez},
  A. 2013, \aap, 560, A109, \dodoi{10.1051/0004-6361/201321397}

\bibitem[{{Juri{\'c}} {et~al.}(2008){Juri{\'c}}, {Ivezi{\'c}}, {Brooks},
  {Lupton}, {Schlegel}, {Finkbeiner}, {Padmanabhan}, {Bond}, {Sesar},
  {Rockosi}, {Knapp}, {Gunn}, {Sumi}, {Schneider}, {Barentine}, {Brewington},
  {Brinkmann}, {Fukugita}, {Harvanek}, {Kleinman}, {Krzesinski}, {Long},
  {Neilsen}, {Nitta}, {Snedden}, \& {York}}]{juric2008}
{Juri{\'c}}, M., {Ivezi{\'c}}, {\v{Z}}., {Brooks}, A., {et~al.} 2008, \apj,
  673, 864, \dodoi{10.1086/523619}

\bibitem[{{Lee} {et~al.}(2011){Lee}, {Beers}, {An}, {Ivezi{\'c}}, {Just},
  {Rockosi}, {Morrison}, {Johnson}, {Sch{\"o}nrich}, {Bird}, {Yanny},
  {Harding}, \& {Rocha-Pinto}}]{lee2011}
{Lee}, Y.~S., {Beers}, T.~C., {An}, D., {et~al.} 2011, \apj, 738, 187,
  \dodoi{10.1088/0004-637X/738/2/187}

\bibitem[{{Lian} {et~al.}(2020){Lian}, {Thomas}, {Maraston}, {Beers}, {Moni
  Bidin}, {Fern{\'a}ndez-Trincado}, {Garc{\'\i}a-Hern{\'a}ndez}, {Lane},
  {Munoz}, {Nitschelm}, {Roman-Lopes}, \& {Zamora}}]{lian2020b}
{Lian}, J., {Thomas}, D., {Maraston}, C., {et~al.} 2020, \mnras, 497, 2371,
  \dodoi{10.1093/mnras/staa2078}

\bibitem[{{Lian} {et~al.}(2022){Lian}, {Zasowski}, {Mackereth}, {Imig},
  {Holtzman}, {Beaton}, {Bird}, {Cunha}, {Fern{\'a}ndez-Trincado}, {Horta},
  {Lane}, {Masters}, {Nitschelm}, \& {Roman-Lopes}}]{lian2022}
{Lian}, J., {Zasowski}, G., {Mackereth}, T., {et~al.} 2022, \mnras, 513, 4130,
  \dodoi{10.1093/mnras/stac1151}

\bibitem[{{Mackereth} {et~al.}(2017){Mackereth}, {Bovy}, {Schiavon},
  {Zasowski}, {Cunha}, {Frinchaboy}, {Garc{\'\i}a Perez}, {Hayden}, {Holtzman},
  {Majewski}, {M{\'e}sz{\'a}ros}, {Nidever}, {Pinsonneault}, \&
  {Shetrone}}]{mackereth2017}
{Mackereth}, J.~T., {Bovy}, J., {Schiavon}, R.~P., {et~al.} 2017, \mnras, 471,
  3057, \dodoi{10.1093/mnras/stx1774}

\bibitem[{{Martig} {et~al.}(2016){Martig}, {Fouesneau}, {Rix}, {Ness},
  {M{\'e}sz{\'a}ros}, {Garc{\'\i}a-Hern{\'a}ndez}, {Pinsonneault}, {Serenelli},
  {Silva Aguirre}, \& {Zamora}}]{martig2016}
{Martig}, M., {Fouesneau}, M., {Rix}, H.-W., {et~al.} 2016, \mnras, 456, 3655,
  \dodoi{10.1093/mnras/stv2830}

\bibitem[{{Meng} \& {Gnedin}(2021)}]{meng2021}
{Meng}, X., \& {Gnedin}, O.~Y. 2021, \mnras, 502, 1433,
  \dodoi{10.1093/mnras/stab088}

\bibitem[{{Nidever} {et~al.}(2014){Nidever}, {Bovy}, {Bird}, {Andrews},
  {Hayden}, {Holtzman}, {Majewski}, {Smith}, {Robin}, {Garc{\'\i}a P{\'e}rez},
  {Cunha}, {Allende Prieto}, {Zasowski}, {Schiavon}, {Johnson}, {Weinberg},
  {Feuillet}, {Schneider}, {Shetrone}, {Sobeck}, {Garc{\'\i}a-Hern{\'a}ndez},
  {Zamora}, {Rix}, {Beers}, {Wilson}, {O'Connell}, {Minchev}, {Chiappini},
  {Anders}, {Bizyaev}, {Brewington}, {Ebelke}, {Frinchaboy}, {Ge}, {Kinemuchi},
  {Malanushenko}, {Malanushenko}, {Marchante}, {M{\'e}sz{\'a}ros}, {Oravetz},
  {Pan}, {Simmons}, \& {Skrutskie}}]{nidever2014}
{Nidever}, D.~L., {Bovy}, J., {Bird}, J.~C., {et~al.} 2014, \apj, 796, 38,
  \dodoi{10.1088/0004-637X/796/1/38}

\bibitem[{{Perrin} {et~al.}(2014){Perrin}, {Sivaramakrishnan}, {Lajoie},
  {Elliott}, {Pueyo}, {Ravindranath}, \& {Albert}}]{perrin2014}
{Perrin}, M.~D., {Sivaramakrishnan}, A., {Lajoie}, C.-P., {et~al.} 2014, in
  Society of Photo-Optical Instrumentation Engineers (SPIE) Conference Series,
  Vol. 9143, Space Telescopes and Instrumentation 2014: Optical, Infrared, and
  Millimeter Wave, ed. J.~{Oschmann}, Jacobus~M., M.~{Clampin}, G.~G. {Fazio},
  \& H.~A. {MacEwen}, 91433X, \dodoi{10.1117/12.2056689}

\bibitem[{{Pinna} {et~al.}(2019){Pinna}, {Falc{\'o}n-Barroso}, {Martig},
  {Sarzi}, {Coccato}, {Iodice}, {Corsini}, {de Zeeuw}, {Gadotti}, {Leaman},
  {Lyubenova}, {McDermid}, {Minchev}, {Morelli}, {van de Ven}, \&
  {Viaene}}]{pinna2019}
{Pinna}, F., {Falc{\'o}n-Barroso}, J., {Martig}, M., {et~al.} 2019, \aap, 623,
  A19, \dodoi{10.1051/0004-6361/201833193}

\bibitem[{{Queiroz} {et~al.}(2020){Queiroz}, {Anders}, {Chiappini},
  {Khalatyan}, {Santiago}, {Steinmetz}, {Valentini}, {Miglio}, {Bossini},
  {Barbuy}, {Minchev}, {Minniti}, {Garc{\'\i}a Hern{\'a}ndez}, {Schultheis},
  {Beaton}, {Beers}, {Bizyaev}, {Brownstein}, {Cunha},
  {Fern{\'a}ndez-Trincado}, {Frinchaboy}, {Lane}, {Majewski}, {Nataf},
  {Nitschelm}, {Pan}, {Roman-Lopes}, {Sobeck}, {Stringfellow}, \&
  {Zamora}}]{queiroz2020}
{Queiroz}, A.~B.~A., {Anders}, F., {Chiappini}, C., {et~al.} 2020, \aap, 638,
  A76, \dodoi{10.1051/0004-6361/201937364}

\bibitem[{{Quinn} {et~al.}(1993){Quinn}, {Hernquist}, \&
  {Fullagar}}]{quinn1993}
{Quinn}, P.~J., {Hernquist}, L., \& {Fullagar}, D.~P. 1993, \apj, 403, 74,
  \dodoi{10.1086/172184}

\bibitem[{{Reddy} {et~al.}(2006){Reddy}, {Lambert}, \& {Allende
  Prieto}}]{reddy2006}
{Reddy}, B.~E., {Lambert}, D.~L., \& {Allende Prieto}, C. 2006, \mnras, 367,
  1329, \dodoi{10.1111/j.1365-2966.2006.10148.x}

\bibitem[{{Sandin}(2014)}]{sandin2014}
{Sandin}, C. 2014, \aap, 567, A97, \dodoi{10.1051/0004-6361/201423429}

\bibitem[{{Shapiro} {et~al.}(2008){Shapiro}, {Genzel}, {F{\"o}rster Schreiber},
  {Tacconi}, {Bouch{\'e}}, {Cresci}, {Davies}, {Eisenhauer}, {Johansson},
  {Krajnovi{\'c}}, {Lutz}, {Naab}, {Arimoto}, {Arribas}, {Cimatti}, {Colina},
  {Daddi}, {Daigle}, {Erb}, {Hernandez}, {Kong}, {Mignoli}, {Onodera},
  {Renzini}, {Shapley}, \& {Steidel}}]{shapiro2008}
{Shapiro}, K.~L., {Genzel}, R., {F{\"o}rster Schreiber}, N.~M., {et~al.} 2008,
  \apj, 682, 231, \dodoi{10.1086/587133}

\bibitem[{{Simons} {et~al.}(2016){Simons}, {Kassin}, {Trump}, {Weiner},
  {Heckman}, {Barro}, {Koo}, {Guo}, {Pacifici}, {Koekemoer}, \&
  {Stephens}}]{simons2016}
{Simons}, R.~C., {Kassin}, S.~A., {Trump}, J.~R., {et~al.} 2016, \apj, 830, 14,
  \dodoi{10.3847/0004-637X/830/1/14}

\bibitem[{{Stefanon} {et~al.}(2017){Stefanon}, {Yan}, {Mobasher}, {Barro},
  {Donley}, {Fontana}, {Hemmati}, {Koekemoer}, {Lee}, {Lee}, {Nayyeri}, {Peth},
  {Pforr}, {Salvato}, {Wiklind}, {Wuyts}, {Ashby}, {Castellano}, {Conselice},
  {Cooper}, {Cooray}, {Dolch}, {Ferguson}, {Galametz}, {Giavalisco}, {Guo},
  {Willner}, {Dickinson}, {Faber}, {Fazio}, {Gardner}, {Gawiser}, {Grazian},
  {Grogin}, {Kocevski}, {Koo}, {Lee}, {Lucas}, {McGrath}, {Nandra}, {Newman},
  \& {van der Wel}}]{stefanon2017}
{Stefanon}, M., {Yan}, H., {Mobasher}, B., {et~al.} 2017, \apjs, 229, 32,
  \dodoi{10.3847/1538-4365/aa66cb}

\bibitem[{{Tsikoudi}(1979)}]{tsikoudi1979}
{Tsikoudi}, V. 1979, \apj, 234, 842, \dodoi{10.1086/157565}

\bibitem[{{{\"U}bler} {et~al.}(2019){{\"U}bler}, {Genzel}, {Wisnioski},
  {F{\"o}rster Schreiber}, {Shimizu}, {Price}, {Tacconi}, {Belli}, {Wilman},
  {Fossati}, {Mendel}, {Davies}, {Beifiori}, {Bender}, {Brammer}, {Burkert},
  {Chan}, {Davies}, {Fabricius}, {Galametz}, {Herrera-Camus}, {Lang}, {Lutz},
  {Momcheva}, {Naab}, {Nelson}, {Saglia}, {Tadaki}, {van Dokkum}, \&
  {Wuyts}}]{ubler2019}
{{\"U}bler}, H., {Genzel}, R., {Wisnioski}, E., {et~al.} 2019, \apj, 880, 48,
  \dodoi{10.3847/1538-4357/ab27cc}

\bibitem[{{Valentino} {et~al.}(2023){Valentino}, {Brammer}, {Gould}, {Kokorev},
  {Fujimoto}, {Jespersen}, {Vijayan}, {Weaver}, {Ito}, {Tanaka}, {Ilbert},
  {Magdis}, {Whitaker}, {Faisst}, {Gallazzi}, {Gillman}, {Gim{\'e}nez-Arteaga},
  {G{\'o}mez-Guijarro}, {Kubo}, {Heintz}, {Hirschmann}, {Oesch}, {Onodera},
  {Rizzo}, {Lee}, {Strait}, \& {Toft}}]{valentino2023}
{Valentino}, F., {Brammer}, G., {Gould}, K. M.~L., {et~al.} 2023, \apj, 947,
  20, \dodoi{10.3847/1538-4357/acbefa}

\bibitem[{{Villalobos} \& {Helmi}(2008)}]{villalobos2008}
{Villalobos}, {\'A}., \& {Helmi}, A. 2008, \mnras, 391, 1806,
  \dodoi{10.1111/j.1365-2966.2008.13979.x}

\bibitem[{{Villumsen}(1985)}]{villumsen1985}
{Villumsen}, J.~V. 1985, \apj, 290, 75, \dodoi{10.1086/162960}

\bibitem[{{Weiner} {et~al.}(2006){Weiner}, {Willmer}, {Faber}, {Melbourne},
  {Kassin}, {Phillips}, {Harker}, {Metevier}, {Vogt}, \& {Koo}}]{weiner2006}
{Weiner}, B.~J., {Willmer}, C. N.~A., {Faber}, S.~M., {et~al.} 2006, \apj, 653,
  1027, \dodoi{10.1086/508921}

\bibitem[{{Wyse} {et~al.}(2006){Wyse}, {Gilmore}, {Norris}, {Wilkinson},
  {Kleyna}, {Koch}, {Evans}, \& {Grebel}}]{wyse2006}
{Wyse}, R. F.~G., {Gilmore}, G., {Norris}, J.~E., {et~al.} 2006, \apjl, 639,
  L13, \dodoi{10.1086/501228}

\bibitem[{{Xiang} {et~al.}(2018){Xiang}, {Shi}, {Liu}, {Yuan}, {Chen}, {Huang},
  {Wang}, {Wu}, {Tian}, {Huo}, {Zhang}, \& {Zhang}}]{xiang2018}
{Xiang}, M., {Shi}, J., {Liu}, X., {et~al.} 2018, \apjs, 237, 33,
  \dodoi{10.3847/1538-4365/aad237}

\bibitem[{{Yi} {et~al.}(2023){Yi}, {Jang}, {Devriendt}, {Dubois}, {Han},
  {Kimm}, {Kraljic}, {Park}, {Peirani}, {Pichon}, \& {Rhee}}]{yi2023}
{Yi}, S.~K., {Jang}, J.~K., {Devriendt}, J., {et~al.} 2023, arXiv e-prints,
  arXiv:2308.03566, \dodoi{10.48550/arXiv.2308.03566}

\bibitem[{{Yoachim} \& {Dalcanton}(2006)}]{yoachim2006}
{Yoachim}, P., \& {Dalcanton}, J.~J. 2006, \aj, 131, 226,
  \dodoi{10.1086/497970}

\bibitem[{{Yu} {et~al.}(2021){Yu}, {Li}, {Chen}, {Huang}, {Jia}, {Xiang},
  {Yuan}, {Shi}, {Wang}, \& {Liu}}]{yu2021}
{Yu}, Z., {Li}, J., {Chen}, B., {et~al.} 2021, \apj, 912, 106,
  \dodoi{10.3847/1538-4357/abf098}

\end{thebibliography}

\end{document}